\documentclass[aps,10pt,prl,twocolumn,amsmath,amssymb,superscriptaddress]{revtex4-1}

\usepackage[english]{babel} 
\usepackage[utf8]{inputenc}
\usepackage[T1]{fontenc}

\usepackage{lipsum, babel}

\usepackage{bbm}
\usepackage{verbatim}
\usepackage{graphicx}
\usepackage{times}
\usepackage{epsfig}
\usepackage{bm}
\usepackage{txfonts}
\usepackage{dsfont}
\usepackage{color}

\usepackage{hyperref}
\hypersetup{
colorlinks=true,
linkcolor=blue,          % color of internal links
citecolor=blue,          % color of links to bibliography
filecolor=magenta,       % color of file links
urlcolor=red
}

\newcommand{\ket}[1]{|#1\rangle}
\newcommand{\bra}[1]{\langle#1|}

\begin{document}

\selectlanguage{english}

\title{
Quantum Synchronization Blockade:\\ 
Energy Quantization hinders Synchronization of Identical Oscillators
}

\author{Niels L\"orch }
\affiliation{Department of Physics, University of Basel, Klingelbergstrasse 82, CH-4056 Basel, Switzerland}

\author{Simon E. Nigg}
\affiliation{Department of Physics, University of Basel, Klingelbergstrasse 82, CH-4056 Basel, Switzerland}

\author{Andreas Nunnenkamp}
\affiliation{Cavendish Laboratory, University of Cambridge, Cambridge CB3 0HE, United Kingdom}

\author{Rakesh P. Tiwari}
\affiliation{Department of Physics, University of Basel, Klingelbergstrasse 82, CH-4056 Basel, Switzerland}
\affiliation{Department of Physics, McGill University, Montreal, Quebec, Canada}

\author{Christoph Bruder}
\affiliation{Department of Physics, University of Basel, Klingelbergstrasse 82, CH-4056 Basel, Switzerland}

\date{\today}

\begin{abstract}
  Classically, the tendency towards spontaneous synchronization is
  strongest if the natural frequencies of the self-oscillators are as
  close as possible.  We show that this wisdom fails in
  the deep quantum regime, where the uncertainty of amplitude narrows
  down to the level of single quanta. Under these circumstances
  identical self-oscillators cannot synchronize and detuning their
  frequencies can actually {\it help} synchronization.  The effect can
  be understood in a simple picture: Interaction requires an exchange
  of energy. In the quantum regime, the possible quanta of energy are
  discrete. If the extractable energy of one oscillator does not
  exactly match the amount the second oscillator may absorb,
  interaction, and thereby synchronization is blocked.  We demonstrate
  this effect, which we coin \textit{quantum synchronization
    blockade}, in the minimal example of two Kerr-type
  self-oscillators and predict consequences for small oscillator
  networks, where synchronization between blocked oscillators can be
  mediated via a detuned oscillator. We also propose concrete
  implementations with superconducting circuits and trapped ions. This
  paves the way for investigations of new quantum synchronization phenomena in
  oscillator networks both theoretically and experimentally.
\end{abstract}


\maketitle

Coupled self-oscillating systems can spontaneously synchronize, i.e.,
align their phase and frequency.  This phenomenon \cite{Pikovsky2003, Balanov2009} is observed in a
multitude of systems, ranging from the spontaneous blinking of
fireflies in unison to the firing of neurons in the human brain, and technical applications  such as lasers.

The laser is a well-known example of a quantum system that is described as a self-osillator. However, its steady state far above threshold settles into a coherent state,
which is essentially classical \cite{Gardiner2004b, Walls2008}.
Therefore, its synchronization behavior could so far be fully described
within a semiclassical picture \cite{Schleich1988a, Schleich1988}, which allows for efficient simulations.
Along this line, powerful methods have been developed capable of describing large quantum oscillator arrays, such as complex lasing media \cite{Tureci2006, Malik2015}, arrays of optomechanical systems \cite{Ludwig2013, Weiss2015} and polariton condensates \cite{Eastham2008, Wouters2008a, Khan2016, Sieberer2016}.

The rapid experimental progress \cite{Leibfried2003,Koch2007,Schoelkopf2008,Dykman2012} in the control of
quantum oscillators and in the engineering of their dissipative reservoirs \cite{Poyatos1996, Myatt2000, Vahala2009, Kienzler2015, Holland2015, Rossnagel2016, Fitzpatrick2017} is opening the opportunity to study synchronization
deep in the quantum regime, where only a few energy states are populated \cite{Fazio2013,  Lee2013, Lee2014a, Walter2014a, Walter2014, Xu2014b, Xu2015,Fazio2015, Lorch2016, Weiss2016, Galve2016, Giorgi2016}. 
In this regime semiclassical methods can fail \cite{Lee2013} and anharmonicity on the level of single quanta has been identified \cite{Lorch2016} as a crucial ingredient to demonstrate quantum effects in synchronization.

\begin{figure}[htb]
 \centering
   \includegraphics[width=0.45\textwidth]{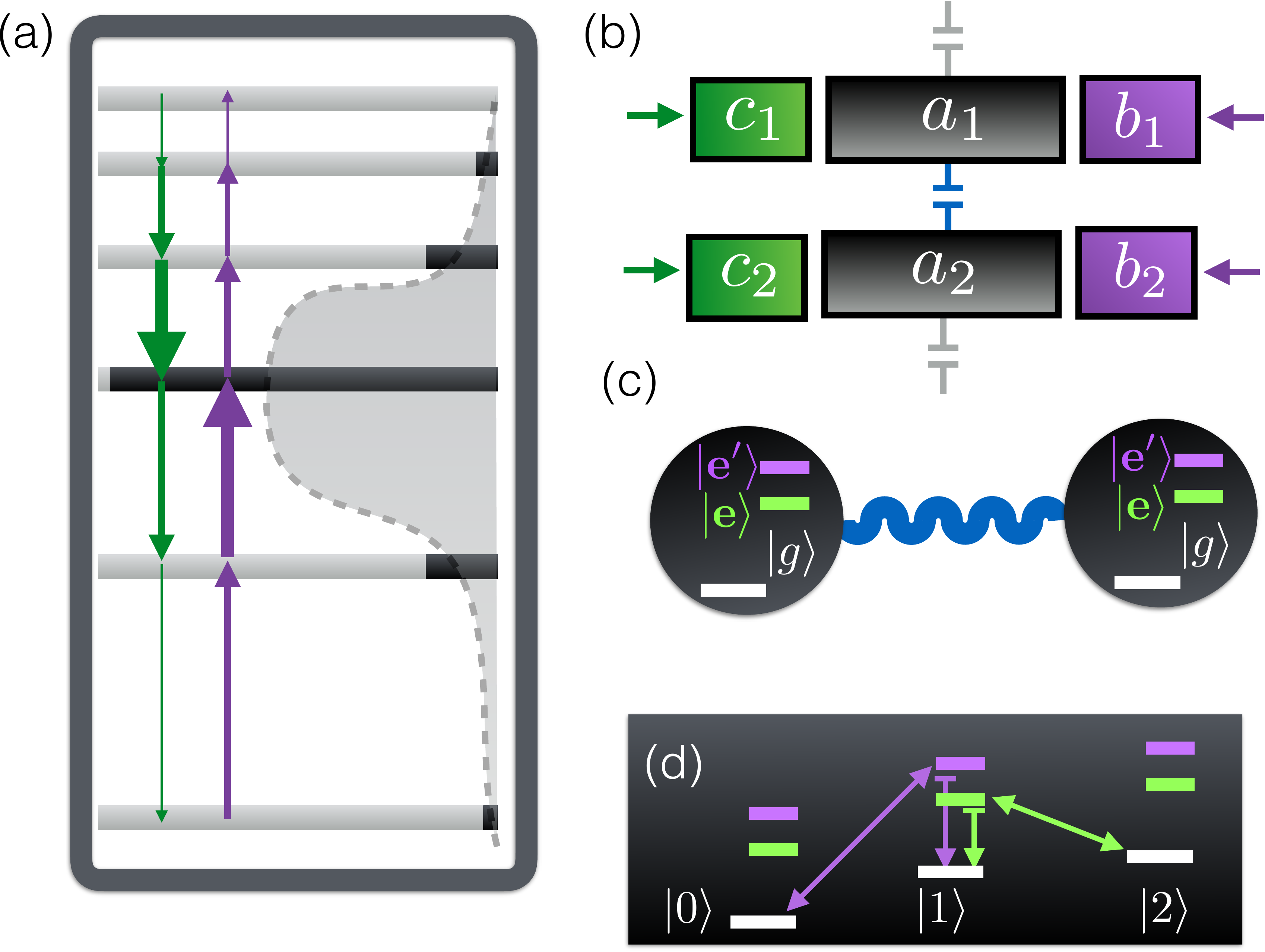}
   \caption{(a) Illustration of an anharmonic oscillator level
     structure (grey) with nonlinear amplification (purple) and damping
     (green) tuned such that a particular Fock state (here the state
     $\ket{2}$) is stabilized.
     The relative thickness of the arrows indicates the transition rate;
     steady-state population probabilities are depicted in black. The
     corresponding classical system would have a continuous energy
     distribution, which is sketched in grey.
     (b) Implementation with an array of superconducting anharmonic
     oscillators driven by an amplification cavity (purple) and a damping
     cavity (green), see main text.
     (c) Implementation with trapped ions with excited states transition between ground state $\ket g$ and excited states $\ket e$, $\ket {e'}$ enabling respectively sideband cooling and sideband amplification of motion as depicted in (d).
     \label{Schematic}}
\end{figure}

In this paper, we discuss a new class of effects in the
synchronization of quantum self-oscillators:
 For the simplest case of two coupled self-oscillators we find that a
finite frequency detuning between different oscillators may {\it
  enable} synchronization in the quantum regime, while synchronization between (nearly) identical self-oscillators is suppressed.  Relatedly, two identical
oscillators of different amplitude are found to synchronize better
than oscillators of the same amplitude.  These findings are in stark contrast to our classical expectation and elude any semiclassical model.
The effect generalizes to
oscillator networks: identical oscillators, while unable to
synchronize directly, can synchronize via a third detuned
oscillator.
We propose possible implementations
in a network of superconducting
circuits~\cite{Koch2007,Schoelkopf2008, Johansson2014} or using trapped
ions~\cite{Leibfried2003,Home2016} to demonstrate the effect experimentally.  Our study opens up a novel regime of
synchronization with genuine quantum features that can be observed
with state-of-the-art quantum hardware. 

\textit{Quantum Model of the System}.--
We consider a network of anharmonic oscillators each described by the Hamiltonian
\begin{equation}
H = \omega a^\dagger a - K (a^\dagger)^2 a^2
\end{equation}
where $a$ is a bosonic annihilation operator,
$\omega$ is the natural frequency of the oscillator,
and the Kerr parameter $K$ quantifies the anharmonicity.
Crucially, the quantum oscillators are subject to dissipation which drives them into self-sustained oscillations (limit cycles). In the framework of open quantum systems this is modeled with a Lindblad operator
$
\mathcal{L} = \mathcal{L}^{(-)} + \mathcal{L}^{(+)}
$
consisting of damping $ \mathcal{L}^{(-)}$ and amplification  $\mathcal{L}^{(+)}$.

To unravel quantum signatures most clearly, we aim for a narrow
distribution of Fock states in steady state, ideally a single
Fock state. One way to achieve \cite{Rips2012a} this is with highly nonlinear dissipators
\begin{align} \label{Dissip}
&\mathcal{L}^{+}=\frac{\gamma_{+}}2\sum_n f_{+}(n) \mathcal D\left[\sqrt{n} \ket {n} \bra{n-1}\right]\:, \nonumber \\
&\mathcal{L}^{-}=\frac{\gamma_{-}}2\sum_n f_{-}(n) \mathcal D\left[\sqrt{n} \ket {n-1} \bra{n}\right]\:,
\end{align}
where 
the individual terms induce transitions from Fock state $\ket{n}$ to
Fock state $\ket{n-1}$. The transition rates $\propto f_+$ ($f_-$) are highly peaked just below (above) the desired Fock state
\footnote{Note that the dissipators in Eq.~\eqref{Dissip} have no
off-diagonal elements between each other in the Lindblad equation
\cite{Rips2012a}. Depending on the concrete system and parameters
these coherences may be present. The effects we want to show exist
in both cases though, so we use this simpler model here.},
as illustrated in Fig.~\ref{Schematic} (a).
Our physical implementation described below results in
\begin{align}
f_\pm(n)= \frac{\sigma_\pm^2}{{(n-n_\pm)^2 + \sigma_\pm^2}}\:, \label{Lorentzian}
\end{align}
where $n_\pm$ and $\sigma_\pm$ are mean and variance of the Lorentzian.
Choosing $n_+$ near an integer $n_0$ and $n_-$ near $n_0+1$ stabilizes
that particular Fock state $\ket {n_0}$, where a high fidelity is
achieved if both $\sigma_-, \sigma_+ \ll 1$. For simplicity, we choose from here on $\sigma_\pm=\sigma, \gamma_\pm=\gamma$, and $n_-=n_++1$.

This corresponds to the extreme quantum limit of self-oscillations,
where the energy distribution is so sharp that only a single Fock
state is populated. Therefore, due to the phase-number uncertainty,
the phase must be in a superposition of all phases. In comparison, the
state of an ordinary laser, as described by an incoherent mixture of
coherent states, also has an undefined phase of classical uncertainty
but not as a result of superposition.

The quantum master equation for the density matrix $\rho$ of a
complete network of such self-oscillators (numbered with index $j$) that are reactively coupled, is given by
\begin{align}
\label{SYS0}
&\dot \rho = -i \left[\sum_j H_j + V, \rho \right] + \sum_j \mathcal{L}_j \rho,
&V=\sum_{j,k} C_{jk}  a_j^\dagger a_k,
\end{align}
where the commutator between $A$ and $B$ is denoted as $[A , B]$ and the coupling matrix associated with the interaction $V$ fulfills $C_{jk}=C_{jk}^*$ and $C_{jj}=0$.

\textit{Classical Model of the System}.--
We will now introduce the corresponding classical description
to be able to compare the quantum system to its classical limit. The
self-oscillators in Eqs.~\eqref{SYS0} can be described by the Langevin equations 
\begin{align} \label{langev}
\dot \alpha_j = - \left(i \Omega_j(\alpha_j)\alpha_j +  \frac{\Gamma_j(\alpha_j)}{2}  \right) \alpha_j -i C_{jk} \alpha_k + \eta_j
\end{align}
with the classical oscillator amplitudes $\alpha_j$.
Here $\Omega_j(\alpha)= \omega_j - 2 K_j \left| \alpha \right| ^2$  
and $\Gamma_j(\alpha)=\gamma_{j-} f_{j-}(\left| \alpha \right|^2 )
-\gamma_{j-} f_{j+}(\left| \alpha \right|^2 )$ are the
amplitude-dependent frequency and damping rate of the $j$-th oscillator,
and $C_{jk}$ is the coupling matrix from Eq.~\eqref{SYS0}. Finally
$\eta_j$ is a white-noise process with correlator $\langle \eta_k(t)
\eta_j(t') \rangle = \delta_{kj} \delta(t-t') \gamma_T n_T$ where
$n_T$ is the thermal bath occupation and $\gamma_T$ is the coupling rate
to the bath.
For conceptual clarity we adopt here the fully classical picture neglecting quantum noise induced by the damping terms $\propto \gamma_+, \gamma_-$. Our main conclusions are not affected by this choice. A derivation of the semi-classical equations including quantum noise can be found in Ref.~\cite{Grimm2016}.
 
\textit{Synchronization Measures}.--
To quantify synchronization between two oscillators we consider the distribution 
$
P(\phi)=\iint_0^{2 \pi} \mathrm d \phi_1 \mathrm d \phi_2 \delta(\phi_1-\phi_2-\phi) p(\phi_1,\phi_2)
$
of their relative phase $\phi$. For the quantum steady state $\rho_{ss}$, we define  
$p(\phi_1,\phi_2)=\bra{\phi_1, \phi_2} \rho_{ss}  \ket {\phi_1,
  \phi_2}$ with phase states
$\ket \phi=\frac 1 {2 \pi} \sum_{n=0}^\infty e^{i n \phi } \ket n$  \cite{Barak2005}.
For the classical case, we define $p(\phi_1,\phi_2)$ as the
probability of $(\alpha_1, \alpha_2)$ to have phases $(\phi_1,\phi_2)$
in the steady state of Eq.~\eqref{langev}.
In both cases we choose the synchronization measure \cite{Hush2015, davis2016synchronization}
\begin{align} \label{ArmourMeasure}
  S=2\pi\max_\phi[P(\phi)]-1,
\end{align}
i.e. a scaled maximum of the relative phase distribution.

\begin{figure}[t!]
\centering
\includegraphics[width=0.45\textwidth]{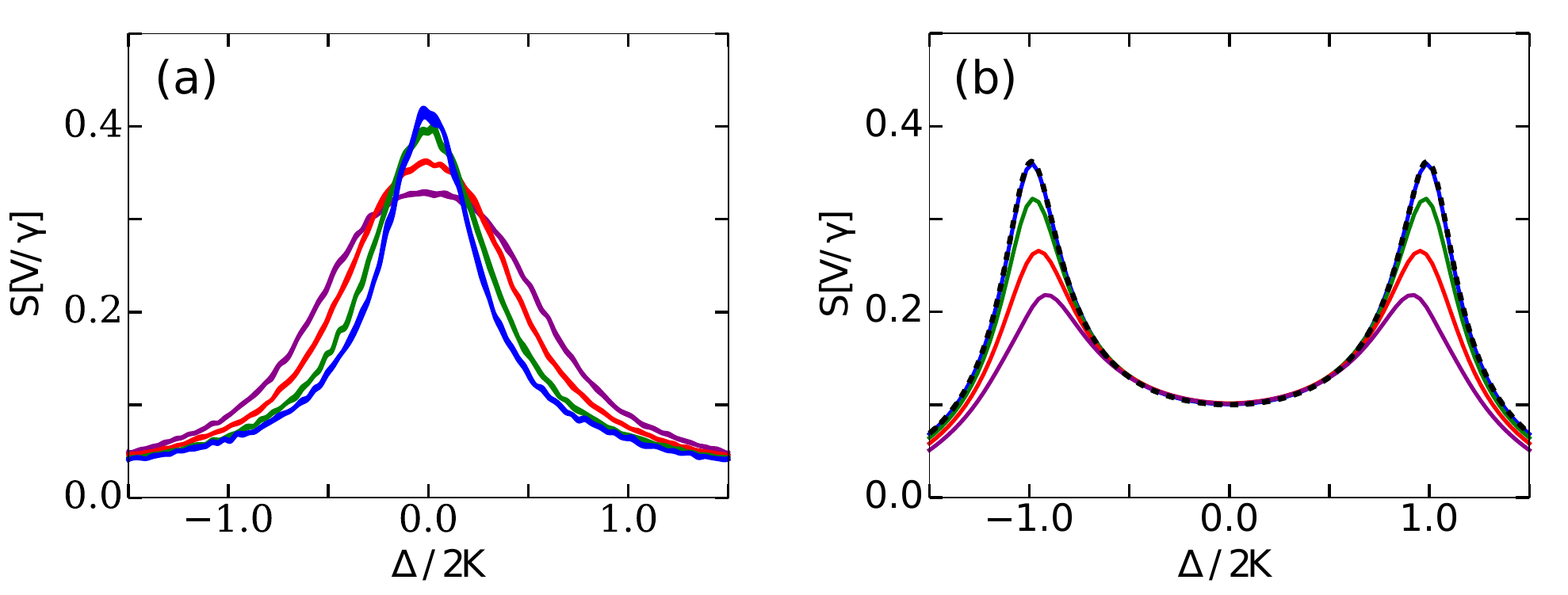}
\caption{ Synchronization measure $S$ calculated using Eq.~\eqref{ArmourMeasure} as a
  function of the detuning $\Delta$ between two oscillators. All other parameters are identical for
  both oscillators. Panel (a) shows the result of classical
  Monte-Carlo simulations of Eq.~\eqref{langev}, where the width of the line indicates the statistical error.
Panel (b) are results from the numerical steady-state solution of the quantum master equation \eqref{SYS0}.  
Classical parameters: $\gamma_T n_t=0.1 \gamma, \sigma=0.2, n_+=2, n_-=3, 
 K=2 \gamma, V=0.1 \gamma \cdot (\frac 16, \frac 14, \frac 38, \frac
 12)$ .
 Quantum parameters: $\sigma=0.2, n_+=2, n_-=3,
 K=10 \gamma, V= \gamma \cdot (0.05, 0.2, 0.35, 0.5)$. In both panels $\gamma_{1\pm}=\gamma_{2\pm}=\gamma$.
  In first-order
 perturbation theory (dashed black
 line), the height of the maximal peak is proportional
 to the coupling $V$ \cite{Supplement}. Plotting  $S$ in units of $V$ for the numerical results, the height of the peak decreases with increasing $V$, where higher-order effects play a role.  
The noise level is chosen such that the effect of the thermal noise in
panel (a) is approximately as strong as the quantum noise in panel (b).
\label{ClassicalSimulation}}
\end{figure}

\textit{Quantum Synchronization Blockade}.--
We now consider the self-oscillator depicted in Fig.~\ref{Schematic} (a)
coupled to another such self-oscillator with all identical parameters,
except for the natural frequencies which are detuned by
$\Delta= \omega_1-\omega_2$.  According to classical intuition, the
strongest tendency to synchronize as a function of $\Delta$ as
measured by \eqref{ArmourMeasure} is always achieved at $\Delta=0$,
where both oscillators are identical. This picture is confirmed by the
numerical solution of Eq.~\eqref{langev}, which is presented in
Fig.~\ref{ClassicalSimulation} (a).  It is consistent with analytical
results obtained in a study of exciton-polariton condensates
\cite{Wouters2008a}, corresponding to the zero-temperature limit of
Eq.~\eqref{langev}.

The classical intuition is not valid in the quantum
system described by Eqs.~\eqref{SYS0}. We investigate the same setup
with parameters deep in the quantum regime, where the limit cycle is
essentially stabilized to a single Fock state $\ket{n_0}$.
The numerical result depicted in Fig.~\ref{ClassicalSimulation} (b)
converge with decreasing coupling strengths to an analytical perturbation theory derived in~\cite{Supplement}.  The phase
synchronization measure is suppressed at $\Delta=0$, where $S$ has a
local minimum. Instead, phase synchronization is now maximal at two
peaks at $\Delta=\pm 2 K$.
 
\begin{figure}[thb]
\centering
\includegraphics[width=0.45\textwidth]{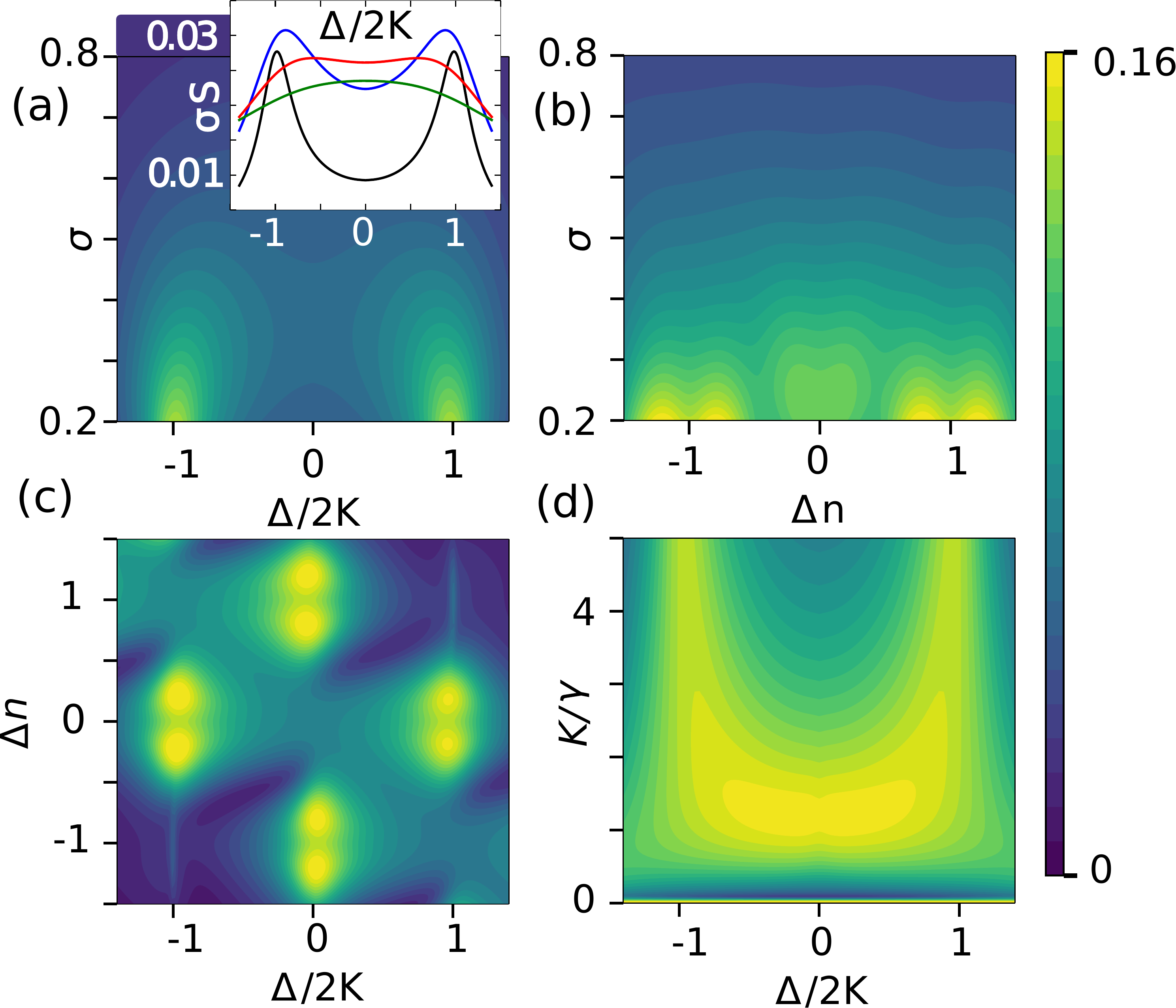}
\caption{Plots of synchronization measure $S$ from Eq.  \eqref{ClassicalSimulation}. (a) Identical oscillators differing only in frequency. Inset shows cuts of $S$ scaled by $\sigma$ at $\sigma=0.2, 0.4, 0.6, 0.8$ in black, blue, red, green. (b) Identical oscillators differing only in amplitude. (c) Overview of these resonances in $\Delta n$ and $\Delta$. (d) Resonances as a function of the Kerr nonlinearity $K$. 
Parameters:
In the upper panels $n_1^+ \equiv 4$ , $n_j^-=n_j^++1$, $V=0.1 \gamma, K=\gamma/\sigma_\pm, \gamma_{1\pm}=\gamma_{2\pm}=\gamma$. In (a) $n_2^+ =4$ and in (b) $\omega_1=\omega_2$. In the lower panels $\sigma_\pm = 0.2$ and all other parameters are as above. 
\label{Overview2}}
\end{figure}

We call this phenomenon the \textit{quantum synchronization blockade},
as it only occurs deep in the quantum regime, where
almost all population is stabilized to a single Fock state.
The transition from quantum to classical is visualized in
Fig.~\ref{Overview2} (a): For a narrow Fock distribution around
$\sigma = 0.2$ the two maxima of synchronization appear at
$\Delta=\pm 2 K$, as just discussed. With increasing width $\sigma$,
the maxima merge to one broad resonance around $\Delta=0$, as classically
expected.
 
In a second scenario we consider self-oscillators of identical
frequency, now differing only in the amplitude $\bar n$ at which they
are stabilized. Oscillator 1 is stabilized to an integer
$\bar n = n_0$ as before, while the amplitude $\bar n$ of oscillator 2
is varied continuously.  The result is shown in Fig.~\ref{Overview2}
(b): In the quantum regime of small $\sigma$ synchronization is
maximal at $\bar n_2=\bar n_1 \pm 1$, i.e. oscillators with a finite difference in amplitude
amplitude are most likely to synchronize.  Again the classical
intuition, that maximal synchronization will be present for identical
oscillators with $\bar n_1=\bar n_2$, is confirmed in the classical
regime of larger $\sigma$.

Thus, in contrast to classical expectation, synchronization of two
quantum oscillators can be enhanced by making the oscillators more
heterogeneous via detuning their frequency or via a mismatch in their
amplitude. The 
result can be explained as follows:
For two oscillators to interact efficiently, the
process $\propto a_k^\dagger a_j$ of exchanging one excitation must be
resonant by conserving energy. For oscillator $j$ in state
$\ket{n_j}$ to transfer an excitation to oscillator $k$ initially in
state $\ket{n_k}$ it is required that 
$E(\ket{n_j,n_k})=E(\ket{n_j-1,n_k+1})$. Writing the energy as
$E(\ket {n_j,n_k})= \bra {n_j,n_k} H \ket {n_j,n_k}=\omega_j n_j +
\omega_k n_k - K (n_j^2+n_k^2-n_j-n_k)$, this leads to the two
resonances
\begin{align} \label{energy condition}
\Delta+2K \Delta n \pm 2K=0\:,
\end{align}
where $\Delta=\omega_j-\omega_k$ and
$\Delta n = n_j-n_k$.
This resonance condition is one of the main results of our paper and
is illustrated in Fig.~\ref{Overview2} (c), showing the synchronization
measure as a function of both $\Delta$ and $\Delta n$ in the quantum
regime. For an illustration of the resonance condition in the case
of identical oscillators, see
 ~\cite{Supplement}.

Equation \eqref{energy condition} includes an offset of $2K$ stemming
from the mismatch of energy in the exchange of a single quantum of
energy described above. Classically, arbitrarily small quanta may be
exchanged, so that the offset does not exist.  For the oscillators to
interact efficiently (and thereby to synchronize), an upward
transition of oscillator $k$ must be resonant with a downward
transition of oscillator $j$, or vice versa.

Fig.~\ref{Overview2} (d) shows how the resonances may be resolved for increasing $K$. For $K=0$ we have the situation of resonant harmonic oscillators ~\cite{Lee2013}, where $P(\phi)$ is a bimodal distribution at $K=0$. Increasing $K$ first leads to a suppression of the resonance and then to a splitting at $\Delta=\pm 2K$. In this regime  only one maximum of $P(\phi)$ survives.

\paragraph*{Oscillator Networks.--} \label{multipleOsc}

\begin{figure}[htb]
\centering
\includegraphics[width=0.45\textwidth]{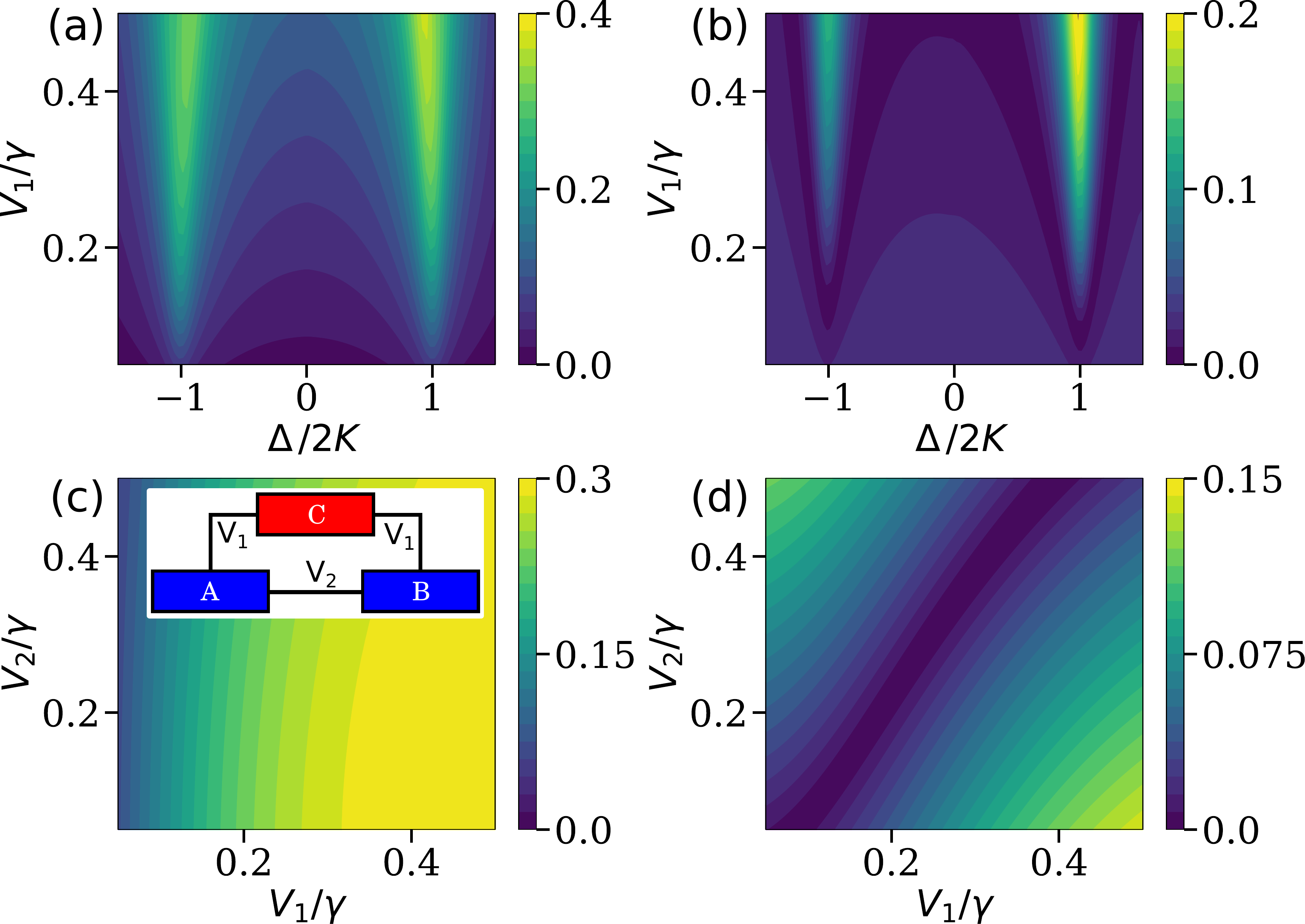}
\caption{Synchronization measure $S$ for the network of three
oscillators depicted in the inset of panel (c). In the upper panels two identical oscillators are connected indirectly via an oscillator $C$ with detuning
$\Delta$, while the direct link $V_2=0$. In the lower panel the direct link $V_2$ is also
turned on  and $\Delta \equiv 2K$.
In both all panels $n_1^+ \equiv 1$, $n_j^-=n_j^++1$, $K=10 \gamma, \gamma_{1\pm}=\gamma_{2\pm}=\gamma$. 
In both rows the left panel shows synchronization
between different oscillators and the right panel between
identical oscillators.
We conclude from the upper panels that
identical oscillators can be synchronized indirectly via a
detuned oscillator.
The lower panels indicate that increasing
the direct coupling can even decrease synchronization for a
strong enough indirect link: In (c) the contour lines bend to the right, while in (d) a strip of suppressed synchronization appears.
\label{Nets}
}
\end{figure}

Having established the quantum synchronization blockade, we now use this understanding to explore
consequences for networks of oscillators. In the following, we focus on
small networks, as these are easiest to implement experimentally.
Consider the three-oscillator network depicted in the inset of Fig.~\ref{Nets} (c), 
where two identical oscillators $A$ and $B$ are coupled indirectly by
connecting them with coupling strength $V_1$ to an oscillator $C$ which has
a relative detuning $\Delta$ 
 with respect to $A$ and $B$. 
 We
first look at the case where the direct coupling $V_2$ between the
two identical oscillators is zero. As shown in Fig.~\ref{Nets} (a),
resonances occur at $\Delta=\pm2K$ for the synchronization between
detuned oscillators, as expected from the two-oscillator
case. Figure~\ref{Nets} (b) shows that identical oscillators can now
synchronize via mediation of the detuned oscillator, again with
resonances at $\Delta=\pm2K$. In this way, the synchronization
blockade can be lifted. This finding is also confirmed for the larger network
of four oscillators, where two pairs of identical oscillators are
connected in a ring in alternating order, see~\cite{Supplement}.

Conversely, as shown in Figs.~\ref{Nets} (c) and (d), turning on a
coupling $V_2$ between the identical oscillators can suppress
synchronization. This effect is most pronounced for identical
oscillators, for which a strip of suppressed synchronization appears
along $V_1 \propto V_2$.

\paragraph*{Implementation.--} 
Nonlinear damping of the form \eqref{Lorentzian} can be
naturally achieved for an anharmonic oscillator mode $a$ coupled to a
linear cavity mode $c$ by coupling the number $\propto c^\dagger c$ of
the cavity to the quadrature $a + a^\dagger$.  Driving the cavity on
the red (blue) sideband will lead to a positive (negative)
damping~\cite{Rips2012a, Grimm2016}. 
Due to the anharmonic level structure the position of the sidebands depends on the oscillator amplitude. Therefore, in contrast to ordinary sideband cooling, the strength of both damping and amplification depends nonlinearly on the oscillation amplitude.

In a rotating frame of the
cavity drive $E$ the Hamiltonian is given by
$ H_c= -\delta c^\dagger c + E(c+c^\dagger) +g_0 c^\dagger c (a +
a^\dagger), $ where $\delta$ is the laser detuning and $g_0$ is the
coupling rate.  Defining
$g=g_0 \sqrt{\langle c^\dagger c \rangle}$ this can be linearized in the regime of large amplitudes (${\langle c^\dagger c \rangle} \gg 1$) as
$
H_c= -\delta c^\dagger c +g (a^\dagger + a) (c + c^\dagger). 
$
Assuming that the cavity decay rate $\kappa$ fulfills $g \ll \kappa$ such that the cavity can be adiabatically eliminated, the parameters of our dissipators \eqref{Dissip} are approximately given by \cite{Rips2012a} $\gamma=4g^2/\kappa$,  $\sigma=\kappa/8K$, and $n_\pm=\pm (\delta_\pm - \omega_0 )/2K$.

Thus, to achieve small $\sigma$ and thereby stabilize a Fock state, a large anharmonicity to cavity noise ratio
$K / \kappa \gtrsim 1$
is required. As depicted in Fig.~\ref{Overview2} (d) and reflected in the perturbation theory
from \cite{Supplement}, also $K / \gamma \gtrsim 1$ is necessary.
 As $\gamma=4g^2/\kappa$ and $g \ll \kappa$, we have the hierarchy  $ \kappa>\gamma$ and therefore only $K / \kappa \gtrsim 1$ remains as the feasibility condition for our specific implementation.
This condition is a challenging requirement
on the experimental setup. For instance, optomechanical systems, while
highly coherent, still lack strong enough anharmonicity.  While this
may be overcome in the future e.g. using auxiliary coupling to a
Cooper pair transistor \cite{Rimberg2014}, we propose an
implementation using superconducting circuits and, alternatively,
trapped ions. In both platforms a large anharmonicity $K \gg \kappa, \gamma$
can be achieved with state-of-the-art technology.

An implementation using superconducting circuits is schematically
depicted in Fig.~\ref{Schematic} (b) for the case of two capacitively
coupled self-oscillators $a_j$. To implement larger networks, the
array can be extended along the greyed out coupling capacitors.  One
choice of self-oscillators are transmon qubits~\cite{Koch2007} which
are sufficiently anharmonic, while offering a long enough coherence
time. The auxiliary cavities for amplification ($b_j$) and damping
($c_j$), are coupled to the self-oscillator via an interaction of
optomechanical form, $c^\dagger c (a+a^\dagger)$. This can be brought
about by embedding a SQUID in the auxiliary
cavity~\cite{Johansson2014}. The particular Lorentzian form
\eqref{Lorentzian} was assumed as a concrete example,
but the scheme is quite general, i.e. any
other setup with both nonlinear damping and amplification 
could be used; any other means
of Fock state stabilization such as
\cite{Holland2015,Souquet2016,Ma2017} will be equally suitable for our
purposes.

An implementation using trapped ions is depicted in
Fig.~\ref{Schematic} (c). Ions trapped in adjacent highly anharmonic
potentials \cite{Home2016} can become self-oscillators  with dissipation engineered as
follows: The roles of the
cavities for amplification and damping are now played by the internal
level structure of the ion, with one transition driven on the blue
sideband and another transition on the red sideband. The use of two transitions is similar
to the schemes  \cite{Lee2013, Hush2015} to implement self-oscillators with ions.  The ions are naturally coupled via the Coulomb
interaction \cite{Hush2015, Brown2011, Harlander2011}.

We note that to observe the effect presented here each node of the network needs to
 have an anharmonic spectrum consisting of at least
three levels, excluding arrays of harmonic oscillators or qubits \cite{Supplement}.

\paragraph{Conclusion and Outlook.--}
\label{Discussion}

To conclude, we have described a novel effect referred to as the
\textit{quantum synchronization blockade}, which prevents identical
nonlinear oscillators from synchronizing deep in the quantum
regime. This is  in stark contrast to the classical regime, where oscillators synchronize best when on resonance. Complementarily we have demonstrated that detuned auxiliary oscillators can lift this blockade by indirectly mediating synchronization between identical oscillators. These effects will be observable in state-of-the-art quantum systems such as superconducting circuits and trapped ions, for which we have proposed concrete implementations.  Our article thus opens a new perspective for the exploration of synchronization in Bose-Hubbard-van der Pol-type networks.

\paragraph*{Acknowledgments.--}
We would like to acknowledge helpful discussions with A. H. Safavi-Naeini and S. M.  Girvin.
This work was financially supported by the Swiss SNF and the NCCR
Quantum Science and Technology. A.N. holds a University Research
Fellowship from the Royal Society and acknowledges support from the
Winton Programme for the Physics of Sustainability. All quantum
steady-state solutions were obtained in QuTiP \cite{Johansson2013} and
the classical Monte-Carlo trajectories were performed in Julia
\cite{Bezanson2014}. This work was supported by the European Union's Horizon 2020 research and innovation programme under grant agreement No 732894 (FET Proactive HOT).

\bibliography{QuantumNetworkSynchronization}

\onecolumngrid

\begin{widetext}

\appendix
\begin{center}
\large\bf Supplemental Material
\end{center}

\section{Analytical perturbation theory in the coupling}
To gain further understanding of the synchronization behavior in the
quantum regime, we rederive the synchronization condition
(7) explicitly based on perturbation theory in
the coupling. To this end we define the perturbation as
$ \mathcal L_1 \rho= -i[V,\rho]$ and the unperturbed Lindbladian as
$\mathcal L_0 \rho=\sum_j -i[H_{0j},\rho] + L_j \rho$, where we use
the definitions from Eq. (4) from the main text and $j=1,2$ indicates the
oscillator. The steady-state density operator to first order in the
coupling $V$ is then
$\rho^{(1)}=-\mathcal L_0^{-1} \mathcal L_1 \rho^{(0)}$, which we
calculate term by term in the following.

The uncoupled density operator $\rho^{(0)}$ is diagonal and for the
individual oscillator its (unnormalized) entries
$ \rho_{mm}^{(0)} \propto (\gamma_+/\gamma_-)^m \prod_{i=1}^mf_+(i)/\prod_{i=1}^mf_-(i)
$
are solved via a recursion relation. The superoperator $\mathcal L_0$ can be
decomposed into a term coupling diagonal density matrix elements and a
term coupling off-diagonal elements separately.
In the off-diagonal subspace, without neglecting any terms, the inverse $\mathcal L_0^{-1}$ can be
obtained by inverting the diagonal so
that\\
$  \mathcal L_0^{-1} \ket{m_1-1,m_2+1} \bra{m_1,m_2} =
  \lambda_{-}^{-1}\ket{m_1-1,m_2+1} \bra{m_1,m_2}
$
with
\begin{align} \label{widths} 
\lambda_{-}&=  -i \left[-\omega_{01}+\omega_{02} + 2K_1 m_1 - 2K_2 m_2- 2K_1
  \right]\nonumber\\
&-\frac{\gamma_{1+}}{2}(m_1 f_{1+}(m_1) + (m_1+1)f_{1+}(m_1+1))-\frac{\gamma_{1-}}{2}((m_1-1)f_{1-}(m_1-1) + m_1f_{1-}(m_1))\nonumber \\
&-\frac{\gamma_{2+}}{2}((m_2+2)f_{2+}(m_2+2) + (m_2+1)f_{2+}(m_2+1))-\frac{\gamma_{2-}}{2}((m_2+1)f_{2-}(m_2+1) + m_2f_{2-}(m_2)) \:. \tag {S1}
\end{align}
As $\mathcal L_1$ couples only neighboring Fock states,
$\rho^{(1)}=-\mathcal L_0^{-1} \mathcal L_1 \rho^{(0)}$ has nonzero
elements only on the minor diagonals, so that the first-order
correction for the steady state is
\begin{align}
\rho^{(1)}_{m_1-1,m_2+1;m_1m_2} = -iC\sqrt{m_1(m_2+1)} \frac{1}{\lambda_-} \left( \rho^{(0)}_{m_1m_2;m_1m_2}-\rho^{(0)}_{m_1-1,m_2+1;m_1-1,m_2+1}\right)
\label{rho_1b} \tag {S2}
\end{align}
and a similar expression for $\rho^{(1)}_{m_1+1,m_2-1;m_1m_2}$.
The synchronization measure (6) from the main text can now be written as the sum
\begin{align} \label{PerturbRes}
   S=2\left|\sum_{m_1=1, m_2=0}^\infty
     \rho^{(1)}_{m_1-1,m_2+1;m_1,m_2}\right| \tag {S3}
\end{align}
of the off-diagonal elements \eqref{rho_1b}.
The analytical expression \eqref{PerturbRes} for $S$ is a sum over
terms \eqref{widths} centered around $2K$ which have widths scaling
with $\gamma_\pm$.  As an approximate rule, we therefore need
$K \gg \gamma_\pm$ to resolve the resonances, which is in agreement
with the numerical simulation shown in Fig. 3 (d) from the main text.

\newpage

\section{Network of four oscillators}

\begin{figure}[htb]
\centering
\includegraphics[width=.8\textwidth]{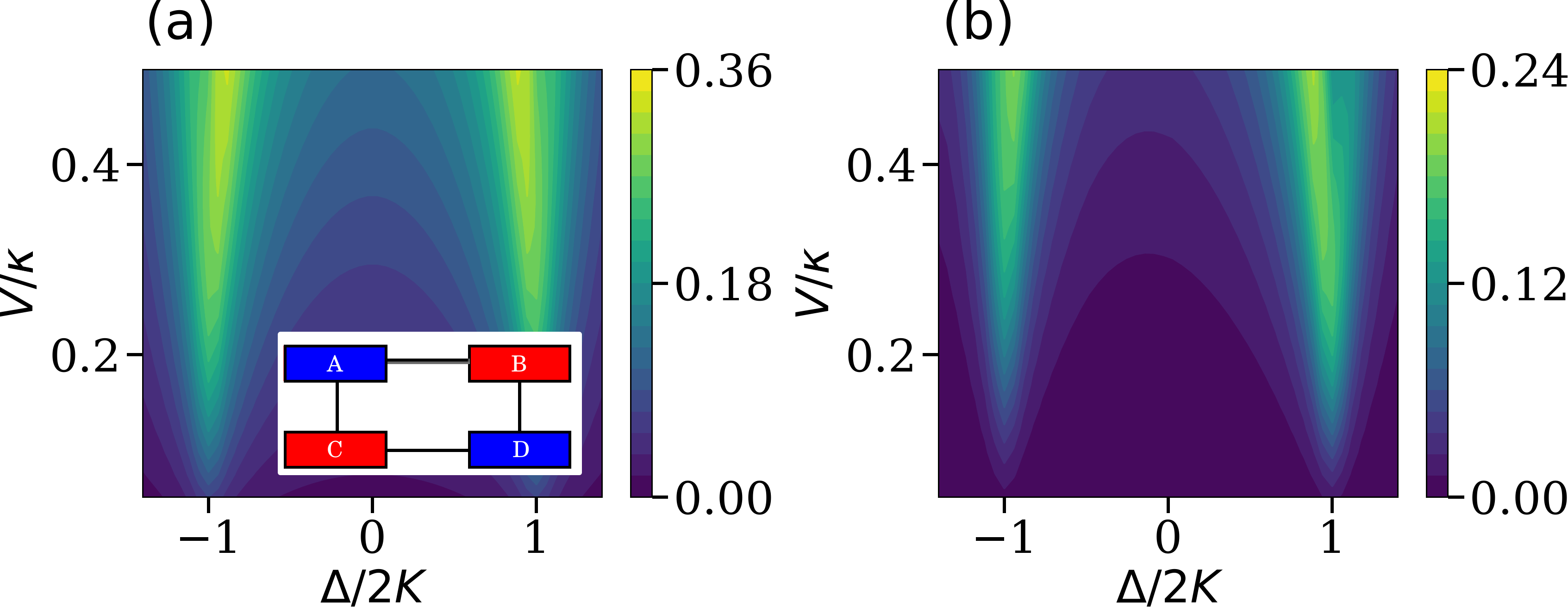}
\caption{Synchronization measure $S$ for the network of four
oscillators depicted in the inset, where neighboring oscillators are coupled with strength $V$. The oscillators $A$ and $D$ are identical and detuned by  by $\Delta$ from another identical set oscillators $B$ and $C$.
All other parameters are as in Fig. 3 of the main text, i.e. $n_1^+ \equiv 1$, $n_j^-=n_j^++1$, $K=10 \gamma, \gamma_{1\pm}=\gamma_{2\pm}=\gamma$. 
Panel (a) shows synchronization as measured by $S$ between a pair of
detuned oscillators, e.g. $A$ and $B$. Panel (b) shows synchronization between identical oscillators, e.g. $A$ and $D$.
The resonances show that the synchronization blockade
between a pair of identical oscillators is lifted by coupling
them via a pair of detuned oscillators. 
\label{Nets4}
}
\end{figure}

We have also considered a network of four oscillators, where two pairs
of identical oscillators are connected in a ring in alternating
order. The two species again differ only in a relative detuning
$\Delta$ of their frequencies $\omega$. As shown in Fig.~\ref{Nets4},
synchronization can be mediated via detuned oscillators, but is
blocked for identical oscillators due to the quantum synchronization
blockade.

\section{Illustration of the effect and minimal model}

In future studies it will be interesting to investigate how the quantum synchronization blockade generalizes to larger networks. It is desirable to restrict the system to a small Hilbert space, to make these simulations numerically feasible. Therefore the individual components of the lattice would ideally be constituted by qubits. However, at least three energy levels per system are required to observe the quantum synchronization blockade effect introduced in this article. 

This can be understood from Figure~\ref{LevelDiagram}, where the origin of the effect is visualized. The crucial diagonal levels are marked in red. These levels are resonant with their neighbors (connected with red lines) for harmonic oscillators. The levels become off-resonant with their neighbors for anharmonic oscillators, resulting in the quantum synchronization blockade for identical oscillators.

A minimal model must keep at least one such red level and its neighbors. The lowest red level with neighbors is $\ket{1, 1}$, which is coupled to $\ket{0, 2}$ and $\ket{2,   0}$. To include these levels, each system must contain at least three levels ($\ket{0}$, $\ket{1}$, $\ket{2}$).
As a model must be anharmonic for the red levels to be off-resonant with their neighbors, an anharmonic oscillator with three levels is the minimum model for the effect to be observable.  The minimal model is further illustrated in Figure~\ref{Levels2}.
It is an interesting direction to investigate whether also spin systems with three or more levels could be a feasible platform for this effect as well. 

The quantum synchronization blockade reported here is unrelated to the observation, that classical anharmonic oscillators with unbalanced losses (resulting in stabilization at different amplitudes) need a finite detuning to compensate for the disbalance to reach maximal synchronization. The suppression of synchronization reported here contradicts the classical expectation, as we have shown. It is also unrelated to the case of two coupled, initially aligned qubits with strongly unbalanced coupling to a thermal quantum bath reported in Ref. 37: Interestingly, the alignment of qubits during time evolution can probe the properties of the bath. Depending on its spectral density, a finite detuning of the unbalanced qubits may favor alignment after transient times before the system equilibrates with the bath.

 \begin{figure}[hb]
  \centering
    \includegraphics[width=.8\textwidth]{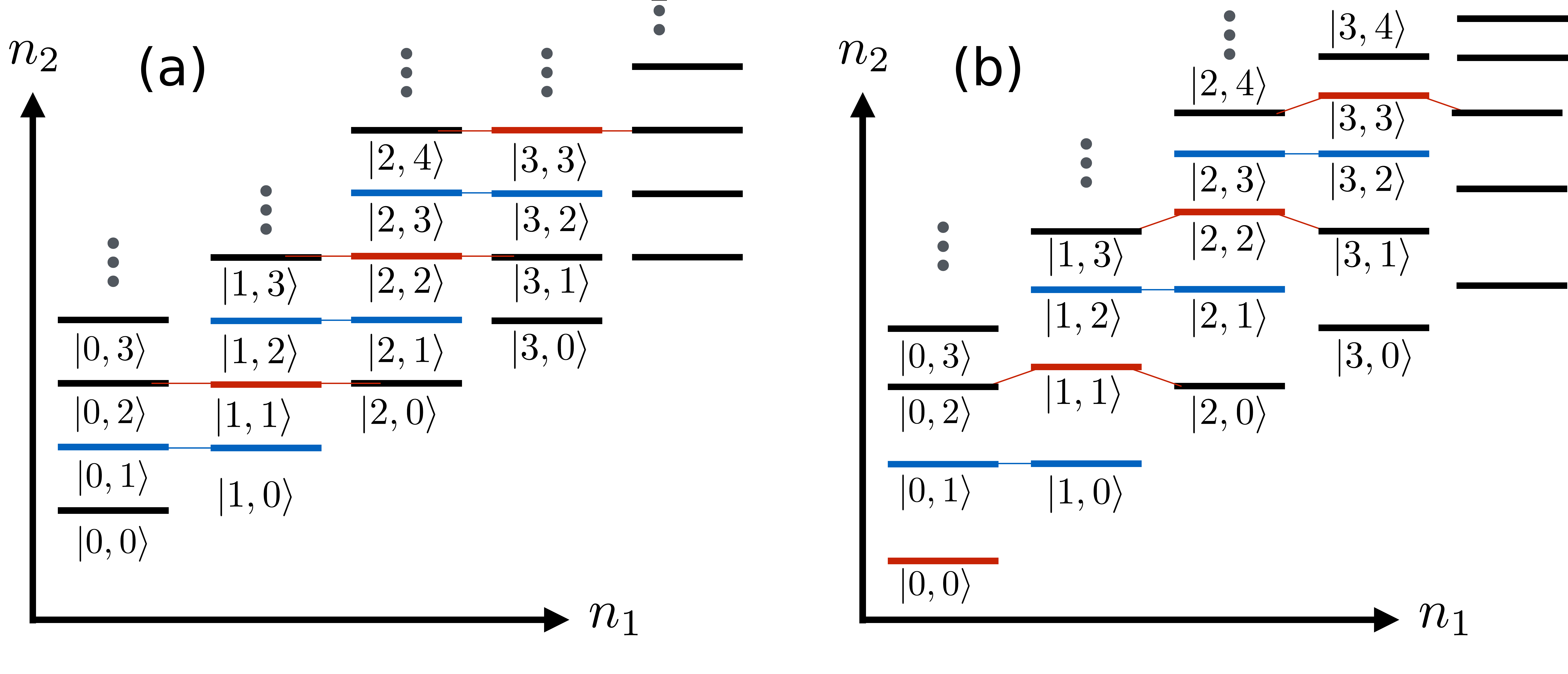}
      \caption{Energy level diagram of two identical oscillators. (a)
        harmonic oscillators. (b) anharmonic oscillators. The axes
        $n_1$ and $n_2$ denote the level of the respective
        oscillators. The red levels, where both oscillators have the
        same occupation, are resonant with the two connected adjacent
        levels in the case of the harmonic oscillator ($K=0$) depicted
        on the left. For the nonlinear oscillator on the right, this
        resonance is lifted. The blue levels corresponding to the
        condition $\Delta n=\pm1$ are resonant in both cases and lead to
        the resonances in Fig. (3) (b) from the main text.
        For example in the column $n_1=2$, the blue levels are
        $\ket {2,1}$ and $\ket {2,3}$, where $\Delta n=\pm1$ is fulfilled.
     \label{LevelDiagram}
   }
\end{figure}

 \begin{figure}[hb]
  \centering
    \includegraphics[width=.9\textwidth]{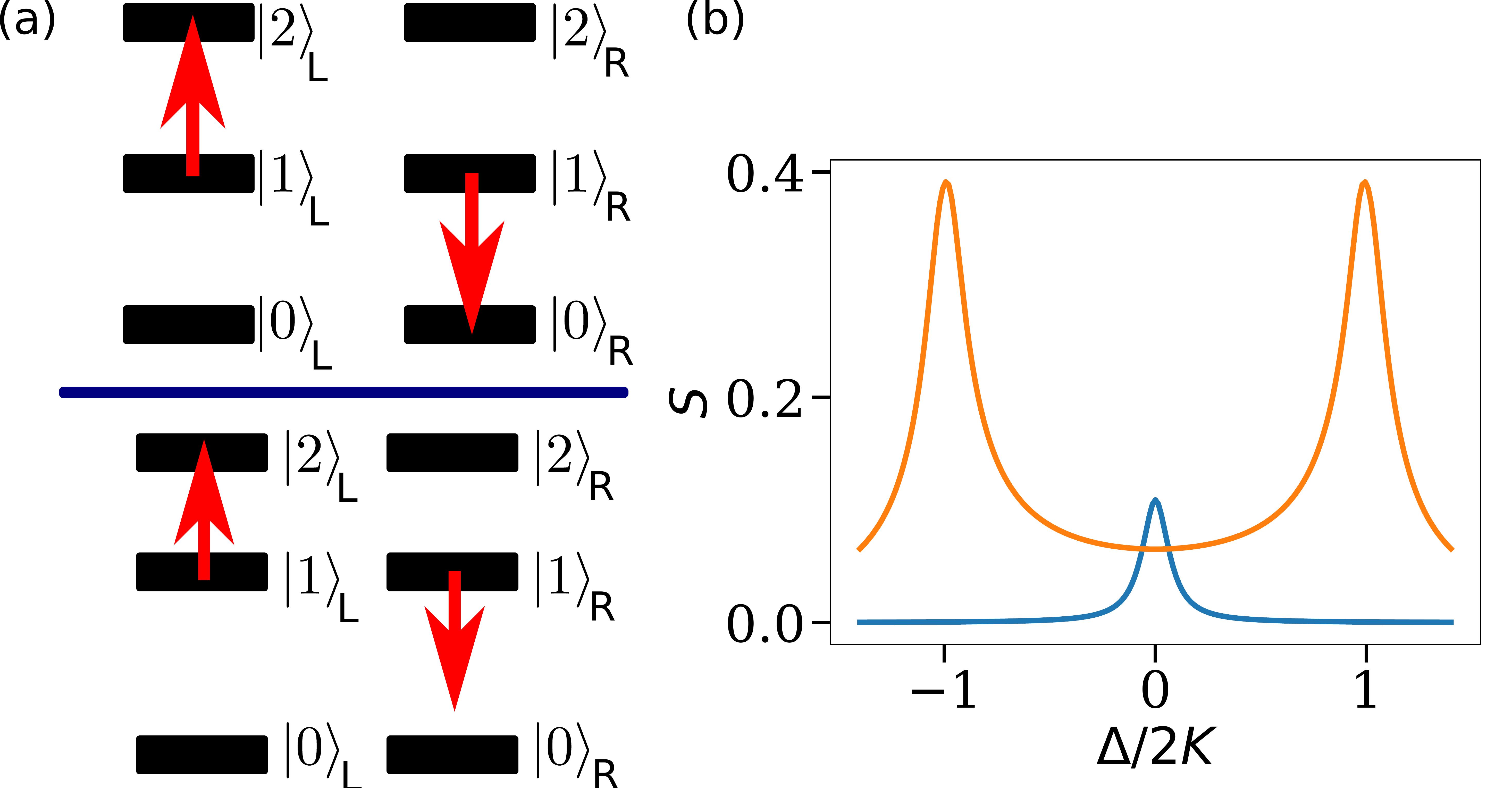}
      \caption{(a) Illustration of the synchronization blockade for the minimal system consisting of two three-level systems with annihilation operators $a_L$ and $a_R$ indexed by $L$ and $R$ for left and right. Before the interaction is turned on, each system is stabilized to level $\ket{1}$. The interaction $a_L^\dagger a_R$  (red arrows) transfers one excitation from system $a_R$ to system $a_L$. For the harmonic case depicted in the upper panel, this conserves energy, as both upward and downward transition have the same energy difference. However for the anharmonic case depicted in the lower panel, the left oscillator gains less energy than the right oscillator loses. Therefore interaction is hindered at zero detuning. With a finite detuning $\Delta=2K$ the process is again resonant. The same argument holds for the reverse process $a_L a_R^\dagger$ at a detuning $\Delta=-2K$. This is the origin of the two side resonances introduced in Fig. 2(b) of the main text.
(b) Synchronization measure $S$ as a function of detuning $\Delta$. For parameters $n_1^+=n_2^+=1, \sigma_1^\pm=\sigma_2^\pm=0.2, V=0.8 \gamma$ the resonance curve of a harmonic oscillator (blue line) with $K=0$ is compared to an anharmonic oscillator (orange line) with $K=3 \gamma/\sigma$. In both cases the oscillators are restricted to the lowest three levels. While the harmonic oscillator confirms the classical expectation of maximal synchronization on resonance, the anharmonic oscillator's resonances are shifted as explained in (a). 
We also numerically checked that in the case of two-level systems $S=0$ for all $\Delta$, as the effect is not applicable to qubits.   }
\label{Levels2}
\end{figure}

\end{widetext}

\end{document}